\documentclass[twocolumn]{aastex61}

\newcommand{\hi}{\ion{H}{1}}

\newcommand{\Msol}{\ensuremath{\mathrm{M_{\odot}}}}
\newcommand{\kms}{\ensuremath{\mathrm{km\,s^{-1}}}}

\newcommand{\Vsol}{\ensuremath{\mathrm{V_{\odot}}}}

\newcommand{\mhi}{\ensuremath{M_{HI}}}
\newcommand{\mstar}{\ensuremath{M_{*}}}
\newcommand{\mhimstar}{\ensuremath{M_{HI}/M_{*}}}

\newcommand{\mdyn}{\ensuremath{M_{dyn}^{3Re}}}
\newcommand{\Wfifty}{\ensuremath{W_{50}}}
\newcommand{\Vrot}{\ensuremath{V_{rot}}}
\newcommand{\Vsys}{\ensuremath{V_{sys}}}
\newcommand{\intS}{\ensuremath{\int{\! \! Sdv}}}
\newcommand{\Vhcg}{\ensuremath{V_{sys,HCG}}}
\newcommand{\rhcg}{\ensuremath{r_{HCG}}}
\newcommand{\Vescr}{\ensuremath{V_{esc}(r)}}
\newcommand{\Veschcg}{\ensuremath{V_{esc}(\rhcg)}}

\newcommand{\gbtidl}{{\sc GBTIDL}}

\newcommand{\vs}{\vspace*{5pt}}

\newcommand{\vegas}{{\sc VEGAS}}

\received{2017 Aug 11}
\revised{2018 Jan 6}
\accepted{XXX}
\submitjournal{ApJ}

\shorttitle{Atomic Gas in Blue UDGs}
\shortauthors{Spekkens \& Karunakaran}

\begin{document}

\title{Atomic Gas in Blue Ultra Diffuse Galaxies around Hickson Compact Groups}

\correspondingauthor{Kristine Spekkens}
\email{kristine.spekkens@gmail.com}

\author{Kristine Spekkens}
\affiliation{Department of Physics\\
Royal Military College of Canada \\
PO Box 17000, Station Forces \\
Kingston, ON K7K 7B4, Canada}
\affiliation{Department of Physics, Engineering Physics and Astronomy \\
Queen's University\\
Kingston, ON K7L 3N6, Canada}

\author{Ananthan Karunakaran}
\affiliation{Department of Physics, Engineering Physics and Astronomy \\
Queen's University\\
Kingston, ON K7L 3N6, Canada}



\begin{abstract}
We have found the atomic gas (\hi) reservoirs of the blue ultra diffuse galaxy (UDG) candidates identified by R\'oman and Trujillo in images near Hickson Compact Groups (HCGs). We confirm that all of the objects are indeed UDGs with effective radii $R_e > 1.5\,$kpc. Three of them are likely to be gravitationally bound to the HCG near which they project, one is plausibly gravitationally bound to the nearest HCG, and one is in the background. We measure \hi\ masses and velocity widths for each object directly from the spectra, and use the widths together with the UDG effective radii to estimate dynamical masses and halo spin parameters. The location of the blue UDGs in the \hi\ mass -- stellar mass plane is consistent with that of the broader gas-rich galaxy population, and both their \hi\ masses and gas richnesses are correlated with their effective radii. The blue UDGs appear to be low-mass objects with high-spin halos, although their properties are not as extreme as those of the faintest diffuse objects found in \hi\ searches. The data presented here highlight the potential of single-dish radio observations for measuring the physical properties of blue diffuse objects detected in the optical.
\end{abstract}




\section{Introduction} \label{sec:intro}

Deep imaging campaigns with both small and large-aperture optical telescopes \citep[e.g.][]{abraham14,trujillo16} have re-invigorated studies of the low surface brightness (LSB) galaxy
population \citep[e.g.][]{impey88,bothun91,oneil00}. Among the more extreme objects revealed are the ultra diffuse galaxies (UDGs), a population of red ($g-i \sim 0.8$), faint ($\mu_g(0) \gtrsim 24 \,\mathrm{mag}\,\mathrm{arc sec}^{-2}$) and extended ($R_e \gtrsim 1.5\,\mathrm{kpc}$) objects in groups and clusters \citep[e.g.][]{vandokkum15,koda15,yagi16,vanderburg16,vanderburg17}.  Since their faintness makes optical spectroscopic follow-up  expensive, very few
distances or dynamical masses for UDGs have been obtained \citep{beasley16,vandokkum16,vandokkum17,kadowaki17}. Nonetheless, a variety of observations suggest that UDGs consist mostly of LSB dwarf
galaxies \citep{beasleytrujillo16,amorisco16,sifon17} with an extreme tail of more massive systems  \citep{vandokkum16,vandokkum17,zaritsky17}.

Optical searches have also revealed populations of blue diffuse objects \citep{james15,roman17,shi17}, while some previously known gas-rich dwarf irregular and LSB galaxies also meet the UDG size and surface brightness criteria \citep{yagi16,bellazzini17,trujillo17}.  Searches for diffuse stellar counterparts to \hi-detected sources have borne fruit as well: \citet{leisman17} identify \hi-bearing ultra diffuse sources (HUDs) in the ALFALFA 70\% catalog \citep{giov05} with similar optical properties to UDGs save for their blue colors. The HUDs appear to be dwarf galaxies embedded in high spin halos. 

There may well be an evolutionary connection between diffuse star-forming field objects and red UDGs, with the former being stripped and quenched upon cluster/group infall to resemble the latter. A variety of models posit that the progenitors of red UDGs represent the high-spin tail of the field galaxy population \citep{yozin15,amorisco16,rong17}. On the other hand, \citet{dicintio17} argue that the progenitors are dwarf galaxies that have undergone multiple episodes of gas outflows from star formation, producing a correlation between effective radius and gas richness in these objects. The physical properties of diffuse blue objects therefore place important constraints on the origin of their red counterparts. At the same time, the gas reservoirs of star-forming low surface brightness objects provide an avenue for measuring distances and internal kinematics through the \hi\ spectral line \citep[e.g.][]{papastergis17}.


\begin{deluxetable}{lccccc}
\tablecaption{Properties of \hi\ spectra  \label{tab:obs}}
\tablewidth{0pt}
\tablehead{
\colhead{Name} & \colhead{$(\alpha,\delta)$} & \colhead{Instr.} &\colhead{$\delta V$} & \colhead{$\sigma_{\delta V}$}  & \colhead{$SNR$} \\
  \colhead{} & \colhead{(J2000 deg)} &  &\colhead{(\kms)} & \colhead{(mJy)} & \colhead{} 
   }
\decimalcolnumbers
\startdata
UDG-B1  & (50.09, -1.17)  & GBT & 10  & 1.2 & 15.2 \\
                &                         & VLA & 21.5 & 0.6 & 9.2\\
UDG-B2 &  (09.60, +1.11) & GBT &10 & 0.6 & 9.0  \\
UDG-B3  & (49.96, -0.86)  & GBT &10 & 0.8 & 10.8 \\
UDG-B4   & (09.89, +1.12) & GBT &20 & 0.4 & 4.9 \\
UDG-B5  &  (09.97, +0.38) & GBT & 20 & 1.4 & 5.8 \\
\enddata
\tablecomments{col.\ (1): name of blue UDG candidate along spectrum LOS, following convention of RT17.  col.\ (2): position of optical feature from RT17.  col. (3): for UDG-B1, source of spectrum. cols.\ (4) and (5): RMS noise $\sigma_{\delta V}$ at spectral resolution $\delta V$ of spectrum along the LOS in col.\ (2) and shown in Fig.~\ref{fig:spectra}. col.\ (6): peak signal to RMS noise of feature in spectrum shown in Fig.~\ref{fig:spectra}.} 
\end{deluxetable}

 \citet[][hereafter RT17]{roman17} search for UDGs in SDSS Stripe 82 imaging near Hickson Compact Groups \citep[HCGs,][]{hickson82}, uncovering nine objects around HCG07 and HCG25 that satisfy the UDG size and surface brightness criteria. Four of them are red and project within $\sim250\,$kpc of the nearest HCG, and the remaining five are blue and project to larger group-centric distances. Passive evolutionary models applied to the blue UDG candidates produce properties that are broadly consistent with the red ones, suggesting a connection between them. A first step in exploring that connection, however, is to confirm a physical association between the blue UDG candidates and HCGs as well as to measure their basic physical properties. 
 
\begin{figure*}
\epsscale{1.1}
\plotone{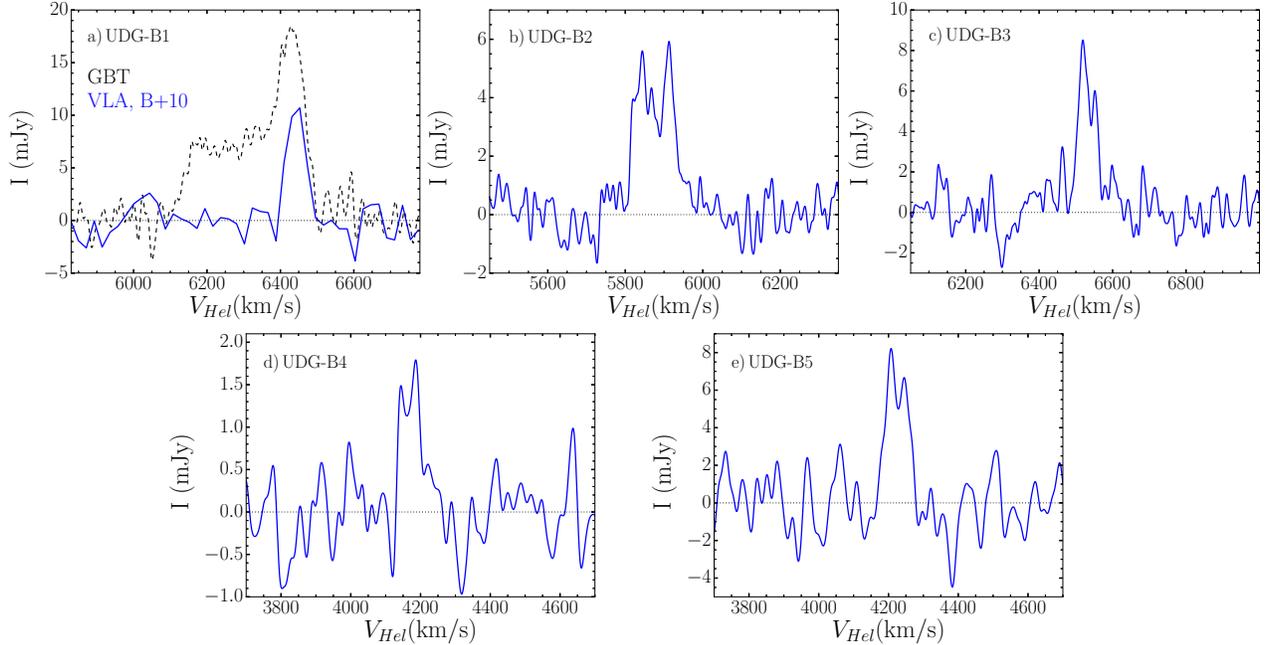}
\caption{\hi\ detections along the line-of-sight to a) UDG-B1, b) UDG-B2, c) UDG-B3, d) UDG-B4, e) UDG-B5. The RMS noise, spectral resolution and signal-to-noise of each spectrum are given in Table~\ref{tab:obs}. For UDG-B1 in panel a), the GBT spectrum is indicated by the dashed black line, and the VLA spectrum is given by the solid blue line. We use the VLA spectrum in our analysis (see text). \label{fig:spectra}}
\end{figure*}

We have detected the \hi\ reservoirs of the blue UDG candidates identified by RT17 with the Robert C. Byrd Green Bank Telescope (GBT), confirming that they are indeed UDGs with $R_e>1.5 \,$kpc. We describe the data acquisition and reduction in \S\ref{sec:obs}. Our methods for deriving distances and \hi\ masses as well as for estimating dynamical masses and halo spin parameters are explained in \S\ref{sec:HIprop}, where we also compare the blue UDG properties to those of other gas-rich galaxy populations. We describe the implications of our measurements for the structure and evolution of UDGs in \S\ref{sec:discuss}.

Throughout, we assume distances to HCG07 and HCG25 of $59\,$Mpc and $88\,$Mpc, respectively, consistent with their recessional velocities and $H_0 = 70\,\kms\,\mathrm{Mpc}^{-1}$.

\section{Observations and Data Reduction} \label{sec:obs}

We obtained five hours of director's discretionary time on the GBT on 2016 December 28-30 (program AGBT16B-424) to carry out position-switched \hi\ observations of the blue UDG candidates reported by RT17. We used the Versatile GBT Astronomical Spectrometer (\vegas) with a bandpass of $100\,$MHz and $3.1\,$kHz channels, tuned to search for \hi\ at heliocentric recessional velocities in the range $1000 \kms \leq \Vsol \leq 21000 \kms$ along the line of sight (LOS) to each UDG candidate and at several reference locations. 

 Integration times (divided equally between signal and reference locations) for each object were determined by predicting its characteristic \hi\ mass \mhi\ and velocity width using the stellar masses \mstar\ computed by RT17 and the scaling relations of \citet{bradford15} for gas-rich field dwarfs. Given our wide bandpass and the distance independence of the ratio \mhi/\mstar\ for a feature of fixed spectral width, our observations are sensitive to gas-rich objects in both the foreground and background of the HCG near which each UDG candidate projects on the sky. The 9.1\arcmin\ full-width at half-maximum (FWHM) of the GBT at $\nu \sim 1.4\,$GHz well exceeds the half-light stellar diameters of the UDG candidates, and we therefore expect any \hi\ therein to be detected by a single pointing. 


The data were reduced using the standard \gbtidl\footnote{ \tt http://gbtidl.nrao.edu/} routine {\it getps}. We then smoothed the data to different resolutions in the range $10 - 50\,\kms$ and examined the resulting spectra for statistically significant emission. We find only one clear \hi\ detection along each LOS. These detections are shown in Fig.~\ref{fig:spectra}, with the RMS noise $\sigma_{\delta V}$ at the resolution $\delta V$ of each spectrum in the figure given in Table~\ref{tab:obs}. We also list the ratio of the peak signal to RMS noise for each feature in Table~\ref{tab:obs}, though its statistical significance is much higher because the emission spans several independent channels. 

The spectrum along the LOS to UDG-B1 is much broader and brighter than that of the other targets. It is also the LOS that projects closest to gas-rich HCG~25, raising the possibility of contamination from its brightest group members. We therefore search for emission at the location of UDG-B1 in the 70\arcsec $\times$50\arcsec -FWHM Very Large Array (VLA) maps of HCG~25 presented by \citet{borthakur10}. We find spatially unresolved \hi\ emission at that location (see Table~\ref{tab:obs}), and the corresponding primary beam-corrected VLA spectrum is presented in Fig.~\ref{fig:spectra}a). Compared to our GBT spectrum, that from the higher angular resolution VLA  is much fainter and narrower. We conclude that the GBT spectrum along that LOS therefore includes a significant contribution from the other HCG~25 group members, and we adopt the VLA spectrum in the analysis of UDG-B1 that follows.

The spectra along the LOS to the blue UDG candidates have shapes characteristic of gas-rich dwarf galaxies, and all but UDG-B1 show hints the double-horned profile expected for a rotating gas-rich disk galaxy \citep[e.g.][]{giov88}; this is not unexpected given the lower spectral resolution of the VLA data (Table~\ref{tab:obs}). Our detections strongly suggest that we have found the \hi\ reservoirs of the blue UDG candidates. For UDG-B1, the spectrum that we extract from the VLA maps is spatially coincident with the optical feature, and its systemic velocity (\S\ref{sec:HIprop:Dindep}) differs from the SDSS DR13 \citep{dr13} spectroscopic redshift by only $10\,\kms$ ($\sim2\sigma$). For UDG-B3, the systemic velocity that we measure is identical within uncertainties to its SDSS DR13 redshift. We searched SDSS imaging and catalogs for potentially gas-rich galaxies within 10\arcmin ($\sim2$ GBT beam radii) of the targeted sky position for the other sources that could produce an \hi\ spectrum consistent with that observed, and find none. We therefore conclude that all of our detections are the \hi\ counterparts to the blue UDG candidates identified by RT17.

  The distances implied by the centroids of the \hi\ spectra in Fig.~\ref{fig:spectra} confirm that they all have effective radii $R_e > 1.5\,$kpc as suggested by RT17 (see \S\ref{sec:HIprop:D}). The candidates are therefore UDGs according to the size criterion of \citet{vandokkum15}, and we henceforth refer to them as such.   


\section{\hi\ properties of the blue UDGs} \label{sec:HIprop}

\begin{deluxetable*}{lcccccccccc}
\tablecaption{Properties of \hi\ Detections \label{tab:params}}
\tablewidth{0pt}
\tablehead{
\colhead{Name} & \colhead{\Vsys} & \colhead{\Wfifty} & \colhead{\intS}  & \colhead{$D$} & \colhead{HCG} & \colhead{$\log(\mhi)$} & \colhead{$\log(M_*)$} & \colhead{$R_e$} & \colhead{\mdyn} & \colhead{$\lambda$} \\
  \colhead{}  & \colhead{(\kms)} & \colhead{(\kms)} & \colhead{(Jy$\,$\kms)} & & \colhead{(Mpc)} & \colhead{($\log[M_\odot]$)} & \colhead{($\log[M_\odot]$)} & \colhead{(kpc)} & \colhead{($10^9\,\Msol$)} & \colhead{}
   }
\decimalcolnumbers
\startdata
UDG-B1$^{a}$  & $6440 \pm {5}$ & ${50} \pm {10}$ & ${0.6} \pm {0.1}$  & 88 &  HCG25 & ${9.1}^{+0.1}_{-0.1}$ & $8.34^{+0.06}_{-0.06}$ & $3.7 \pm 0.4$ & $\sim {3}$ & $\sim {0.2}$ \\
UDG-B3  & $6533 \pm 1$  & $52 \pm 2$  & $0.41 \pm 0.05$ & 88 & HCG25  & $8.88^{+0.06}_{-0.07}$ & $8.53^{+0.06}_{-0.06}$ & $3.2 \pm 0.3$ & $\sim 3$  & $\sim0.2$ \\
\tableline
UDG-B4   & $4168 \pm 3$ & $58 \pm 5$  & $0.09 \pm 0.03$ & 59  &  HCG07 &$7.9^{+0.1}_{-0.2}$ &  $7.85^{+0.06}_{-0.07}$& $1.7 \pm 0.2$ & $\sim 2$ & $\sim0.08$ \\
UDG-B5  & $4229 \pm 3$ & $73 \pm 5$  & $0.5 \pm 0.1$ & 59  &  HCG07 & $8.6^{+0.1}_{-0.2}$ &  $8.34^{+0.06}_{-0.06}$ & $3.1 \pm 0.3$ & $\sim 6$ & $\sim0.1$ \\
\tableline
UDG-B2 & $5873 \pm 1$ & $110 \pm 2$  & $0.54 \pm 0.05$ & 82  & \nodata & $8.93^{+0.05}_{-0.06}$ &  $8.06^{+0.07}_{-0.08}$ & $2.8 \pm 0.3$ & $\sim10$ & $\sim 0.05$ \\
\enddata
\tablecomments{col.\ (2): heliocentric systemic velocity of \hi\ detection. col.\ (3): velocity width at 50\% of the \hi\ profile peak, corrected for cosmological redshift and for instrumental broadening using the relations of \citet{springob05}. col.\ (4): integrated line flux.  col.\ (5): adopted distance to each detection. col.\ (6): Hickson Compact Group to which detection likely belongs. col.\ (7): \hi\ mass, computed using Eq.~\ref{eq:mhi} with \intS\ in col.\ (4) and $D$ in col.\ (5). col.\ (8): stellar mass from RT17, adjusted to $D$ in col.\ (5). col.\ (9): Effective radius from RT17, adjusted to $D$ in col.\ (5).  cols.\ (10) and (11): dynamical mass and spin parameter estimator computed using Eqs.~\ref{eq:mdyn}~and~\ref{eq:lambda}, respectively, with \Wfifty\ in col.\ (3) and $R_e$ in col.\ (9). \\
$a)$ For UDG-BI, \hi\ properties are calculated using the VLA spectrum in Fig.~\ref{fig:spectra}. } 
\end{deluxetable*}

 We compute a variety of properties of the blue UDGs from the \hi\ spectra presented in \S\ref{sec:obs}, and list the results in Table~\ref{tab:params}. We focus first on distance-independent quantities that we derive directly from the profiles (\S\ref{sec:HIprop:Dindep}). We then assign a distance to each candidate by comparing its systemic velocity to that of the nearest HCG (\S\ref{sec:HIprop:D}), and use it to compute distance-dependent properties (\S\ref{sec:HIprop:Ddep}). 

\subsection{Distance-independent quantities \label{sec:HIprop:Dindep} }

\begin{figure}
\epsscale{1.2}
\plotone{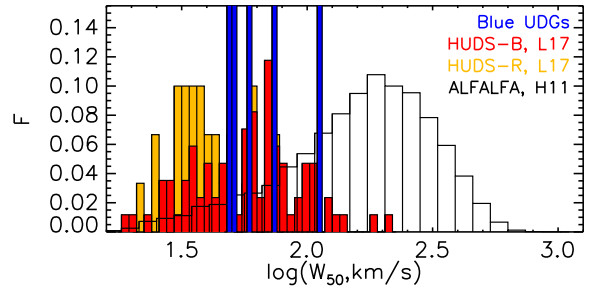}
\caption{Distribution of \Wfifty\ for the blue UDGs (blue histogram), the HUDs-B (red histogram) and HUDs-R (yellow histogram) samples of \hi-bearing ultra diffuse sources from \citet{leisman17}, and code 1 detections in 40\% ALFALFA catalog \citep[unfilled histogram;][]{haynes11}.   \label{fig:wfifty}}
\end{figure}

 Table~\ref{tab:params} presents the heliocentric systemic velocity \Vsys\ and the velocity width \Wfifty\ measured for each of the \hi\ detections.  We follow the method detailed by \citet{springob05} to derive these quantities, fitting a polynomial between 15\% and 85\% of the peak value $f_p$ of each profile edge minus the spectral RMS noise $\sigma_{\delta V}$ from Table~\ref{tab:obs}. We define \Vsys\ to be the mean of the polynomial fit velocities at the 50\% flux level of each edge  ($f=0.5[f_p - \sigma_{\delta V}]$). \Wfifty\ is the difference between these values, corrected for instrumental broadening and cosmological redshift but not for turbulent motions or inclination. As in \citet{springob05}, uncertainties on \Wfifty\ are dominated by that in the instrumental broadening correction, which we take to be 30\% for spectral resolutions $\delta V < 11\,\kms$ and 50\% otherwise. 
 
 Fig.~\ref{fig:wfifty} compares the distribution of \Wfifty\ for the blue UDGs to that of the ``broad" HUDs-B and ``restricted" HUDs-R samples of \hi-bearing ultra-diffuse sources reported by \citet{leisman17}, as well as to code 1 sources in the ALFALFA 40\% catalog \citep{haynes11}. The values of \Wfifty\ for the blue UDGs are generally commensurate with the HUDs-B distribution, and intermediate to the bulk of the HUDs-R sample and the ALFALFA population. This suggests that their dynamical masses are intermediate to typical values for these two samples as well, though a direct physical interpretation is muddled by the lack of disk inclination or gas turbulence corrections to \Wfifty. We estimate dynamical masses for the blue UDGs in \S\ref{sec:HIprop:Ddep}.
 
 
  
 The integrated line flux \intS\ of each detection is also given in Table~\ref{tab:params}, where the uncertainty is dominated by the statistical error in the integral. We use this quantity to compute \hi\ masses in \S\ref{sec:HIprop:Ddep}.
 
 \subsection{Distances \label{sec:HIprop:D} }

With \Vsys\ measurements in-hand, we proceed to assign distances $D$ to each blue UDG. To do this, we compare the measured \Vsys\ to the mean systemic velocity \Vhcg\ of the HCG near which each system projects in order to assess the likelihood that it is gravitationally bound to the group. 

\begin{figure*}
\epsscale{0.8}
\plotone{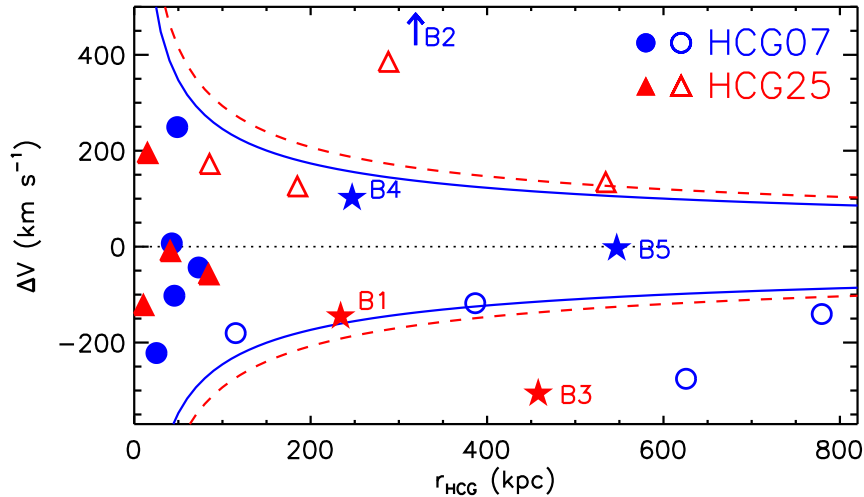}
\caption{Relative velocities as a function of HCG-centric distance for galaxies with $\rhcg < 1\,$Mpc (at the group distance) and $\Delta \Vsys  < 1000\,\kms$ for HCG07 (blue) and HCG25 (red). Blue circles and red triangles show galaxies with measured velocities near HCG07 and HCG25, respectively, and filled symbols show galaxies used to compute the HCG recessional velocity that corresponds to the horizontal black dotted line at $\Delta V = \sqrt{3}(\Vsys - \Vhcg) = 0\,\kms$. The blue solid lines and red dashed lines correspond to the escape velocities of point masses with $M= 1\times 10^{12}\,\Msol$ and $M = 7 \times 10^{11}\,\Msol$ appropriate for HCG07 and HCG25, respectively. The stars show the locations of UDG-B1, UDG-B3, UDG-B4 and UDG-B5, and an upward arrow shows \rhcg\ for B2, whose $\Delta V$ places it well beyond the plot area.   \label{fig:vels}}
\end{figure*}

Fig.~\ref{fig:vels} plots relative recessional velocity as a function of HCG-centric distance \rhcg\ (assuming $D$ to be equal to that of the nearest HCG) for all galaxies with SDSS DR13 spectroscopic redshifts that fall within $\rhcg = 1\,$Mpc and $\Delta \Vsys = 1000\,\kms$ of the sky coordinates and systemic velocities of HCG07 (blue circles) and HCG25 (red triangles). For consistency, we use SDSS DR13 redshifts of confirmed group members \citep[filled circles and triangles;][]{hickson92,konstantopoulos10} to compute \Vhcg\ for each HCG. This value then defines $\Delta V = \sqrt{3}(\Vsys - \Vhcg) = 0\,\kms$ in Fig.~\ref{fig:vels}, where the factor of $\sqrt{3}$ statistically corrects from line-of-sight to 3D space velocities.  

In Fig.~\ref{fig:vels}, the red dashed and blue solid lines correspond to the escape velocities \Vescr\ of point masses with $M= 1\times 10^{12}\,\Msol$ and $M = 7 \times 10^{11}\,\Msol$ as appropriate for the halos of HCG07 and HCG25, respectively \citep[RT17;][]{munari13}. The stars show the locations of UDG-B1 and UDG-B3 (which project near HCG25) as well as UDG-B4 and UDG-B5 (which project near HCG07). The upward arrow shows \rhcg\ for B2 (which projects near HCG07), whose $\Delta V$ places it well above the plot area. 

 Fig.~\ref{fig:vels} illustrates that UDG-B1, UDG-B4 and UDG-B5 all have $| \Delta V | < \Veschcg$, and as such are likely to be gravitationally bound to the nearest HCG. We therefore set $D = D_{HCG25}$ for UDG-B1 and $D=D_{HCG07}$ for UDG-B4 and UDG-B5. On the other hand, $\Delta V >> \Veschcg $ for UDG-B2; it is therefore not physically associated with HCG07. We use the flow model of \citet{mould00} and $H_0 = 70\,\kms \mathrm{Mpc^{-1}}$ to compute a distance $D=82\,$Mpc, which places it in the background of HCG07 by $\sim20\,$Mpc. 
 
 The situation is less clear for UDG-B3: while $ |\Delta V| > \Veschcg$ and $\Delta V < 0$ could imply that it is a foreground object,  it is also true that $\Vsys - \Vhcg \sim \Veschcg$. In other words, UDG-B3 could be gravitationally bound to HCG25 if its line-of-sight velocity is much larger than its tangential velocity relative to the group. We assume that this is the case and set $D=D_{HCG25}$ for UDG-B3, although none of our conclusions change if we instead assign a flow model distance to this system. 
 
 \vspace{20pt}
 
\subsection{Distance-dependent quantities \label{sec:HIprop:Ddep} }

 With the distance to each blue UDG established, we proceed to compute distance-dependent quantities. The \hi\ mass \mhi\ for each system is given by the standard relation
\begin{equation}
\mhi  = 2.356 \times 10^5 D^2 \!\! \intS \,\,\,\,\,\Msol
\label{eq:mhi}
\end{equation}
 for optically thin gas, where $D$ is in Mpc and \intS\ is in $\mathrm{Jy\,\kms}$. We adopt a distance uncertainty of $5\,$Mpc, adding in quadrature with the (dominant) uncertainties on \intS. The resulting values of $\log(\mhi)$ are given in Table~\ref{tab:params}, along with the stellar masses $\log(\mstar)$ reported by RT17 and adjusted to our adopted $D$. 
 
\begin{figure}
\epsscale{1.15}
\plotone{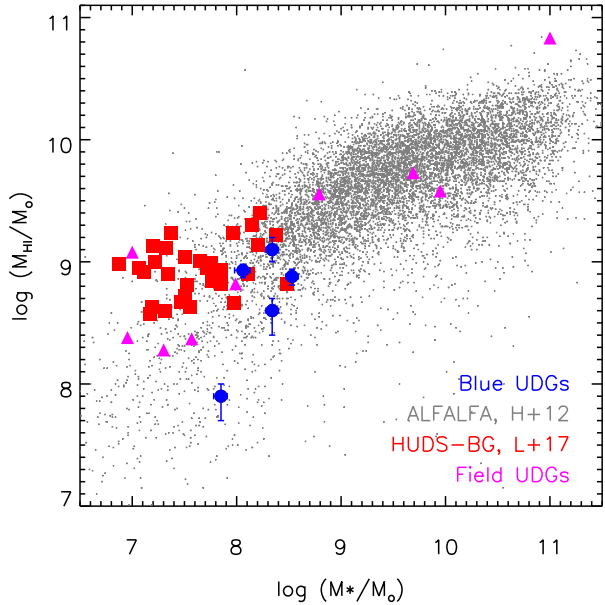}
\caption{Relationship between HI mass and stellar mass for the blue UDGs (blue circles), galaxies in the 40\% ALFALFA catalog with overlapping SDSS and GALEX coverage \citep[][grey points]{huang12b}, the HUDs-BG sample of \hi-bearing ultra diffuse sources from \citet[][red squares]{leisman17}, and the subset of known field UDGs with estimates of \mhi\ and \mstar\  available in the literature.  \label{fig:MHIMstar}}
\end{figure}

 Fig.~\ref{fig:MHIMstar} plots \mhi\ as a function of \mstar\ for the blue UDGs. For comparison, galaxies in the 40\% ALFALFA catalog with overlapping SDSS and GALEX coverage that were analyzed by \citet{huang12b} are shown in grey, galaxies from the \citet{leisman17} HUDs-BG sample (the subset of HUDs-B galaxies that are in the \citealt{huang12b} sample) are shown in red, and the blue field UDGs identified by \citet{yagi16} and \citet{bellazzini17} with estimates of both \mhi\ and \mstar\ in the literature \citep[SdI-1, SdI-2, UGC~2162, UGC~5493, UGC~9024, CGCG~018-057, DDO~87, DDO~143, Malin 1;][]{hunter85,bottinelli90,pickering97,burkholder01,springob05,cook14,chang15,munoz15,papastergis17,trujillo17} are shown in magenta. 
 
 We find that the blue UDGs broadly overlap with the ALFALFA 40\% population in the \mhi\ -- \mstar\ plane, as does the field UDG sample. There is also some overlap between the blue UDGs and the HUDs-BG sample for $\log(\mstar)\gtrsim 7.8$, although the former span a larger range in \mhi\ than the latter. This is in part a selection effect caused by the minimum distance imposed by \citet{leisman17} during their search of the ALFALFA 70\% catalog. 
 
\begin{figure}
\epsscale{1.25}
\plotone{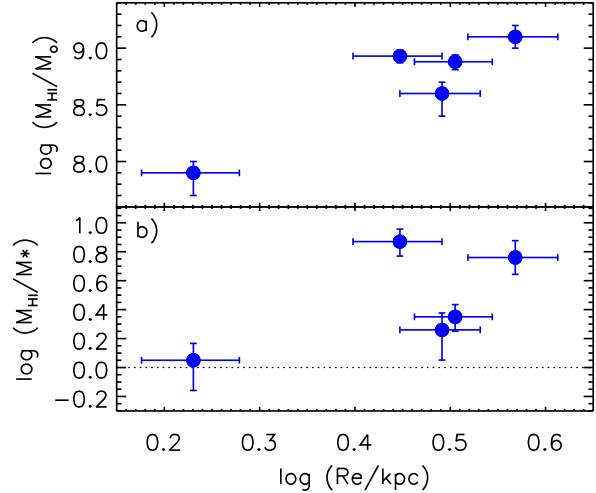}
\caption{Relationship between effective radius and a) \hi\ mass, and b) gas richness \mhimstar\ for blue UDGs. \label{fig:SizeMass}}
\end{figure}

 We list $R_e$ from RT17 for the blue UDGs adjusted to our adopted $D$ in Table~\ref{tab:params}, and plot \mhi\ and the gas richness \mhimstar\ as a function of $R_e$ in Fig.~\ref{fig:SizeMass}. Systems with larger $R_e$ have larger \mhi\ and are more gas-rich, although the correlations are driven largely by the properties of UDG-B4 relative to the other systems.
 

 We use $R_e$ along with \Wfifty\ to estimate dynamical masses for the blue UDGs. Starting from the canonical relation $M = rV^2/G$ for a spherically symmetric system, we assume that the \hi\ distribution extends to $r=3R_e$ \citep{broeils97} and that the rotation velocity at that radius is $V(3R_e) = \Wfifty/(2\sin{i})$, where $i$ is the inclination of the \hi\ disk. For simplicity we assume $i=45^\circ$, which is broadly consistent with the photometric ellipticities measured by RT17. The dynamical mass \mdyn, measured within $3R_e$, is then given by:
\begin{equation}
\mdyn = 3.5\times10^5 R_e \Wfifty^2 \,\,\,\Msol,
\label{eq:mdyn}
\end{equation}
where $R_e$ is measured in kpc and \Wfifty\ is measured in \kms. We emphasize that, given the assumptions regarding the \hi\ disk extent and the relationship between \Wfifty\ and \Vrot\ inherent in Eq.~\ref{eq:mdyn}, \mdyn\ computed for the blue UDGs and given in Table~\ref{tab:params} should be considered as estimates. The dynamical masses of the blue UDGs span a similar range ($\sim0.7\,$dex) to their stellar masses, implying that they are dwarf galaxies.


 To gain some insight into the dark matter halos inside which the UDGs may be embedded, we follow \citet{huang12b} and \citet{leisman17} and compute the halo spin parameter estimator from \citet{hernandez07}.  The estimator assumes that galaxies are embedded in self-gravitating, virialized, isothermal dark matter halos that dominate the potential, and that each one has a flat rotation curve with amplitude \Vrot, an exponential disk with scale-length $R_d$, and a disk mass fraction $\mstar / M_{total} = 0.04$: 
 \begin{equation}
\lambda = 21.8 \frac{R_d}{\Vrot^{3/2}} = 21.8 \frac{R_e}{{\Wfifty}^{3/2}} \,\,\,,
\label{eq:lambda}
\end{equation}
where the second equality results from converting $R_d$ to $R_e$ for an exponential disk and assuming that $\Wfifty= 2 V_{rot} \sin{i}$ with $i = 45^\circ$. While RT17 allow S\'ersic $n$ to vary in their photometric models, the best fitting profiles are almost all consistent with $n=1$ and thus suitable for Eq.~\ref{eq:lambda}. The resulting values of $\lambda$ for the blue UDGs are in Table~\ref{tab:params}. 

We note that even if the numerous assumptions underlying Eq.~\ref{eq:lambda} are correct, the statistical uncertainties on $\lambda$ are large and the values listed in Table~\ref{tab:params} should be interpreted with caution (c.f.\ \citealt{leisman17}). Nonetheless, they enable comparisons with other samples in which $\lambda$ is computed in the same way.  Fig.~\ref{fig:lambda} presents that comparison, plotting $P(\lambda)d \lambda$ for the blue UDGs, that for the ``broad" HUDs-B and ``restricted" HUDs-R samples reported by \citet{leisman17}, as well as the best-fitting lognormal distribution for the ALFALFA 40\% sample computed using Eq.~\ref{eq:lambda} \citep{huang12b}. 

The small number of blue UDGs and the large uncertainties on $\lambda$ imply that standard statistical tests comparing the distributions in Fig.~\ref{fig:lambda} cannot reliably be applied. We instead employ a simpler metric, randomly drawing $10^6$ blue UDG-sized samples from the HUDs-R, HUDs-B and ALFALFA $P(\lambda)d \lambda$ and comparing their mean values to that obtained for the blue UDGs. We find that $>99\%$ of the time, the mean $\lambda$ of a randomly drawn subset of the HUDs-R (ALFALFA) $P(\lambda)d \lambda$ is larger (smaller) than that of the blue UDGs, even when the blue UDG with the least extreme value relative to that distribution is dropped. On the other hand, randomly-drawn subsamples of the HUDs-B $P(\lambda)d \lambda$ have larger mean values than the blue UDG sample $\sim40\%$  of the time. This suggests that $\lambda$ for the blue UDGs are more extreme than those estimated for ALFALFA but less extreme than those from the HUDs-R sample. Instead, they are consistent with being drawn from the HUDs-B sample.

\vs
\section{Discussion and Conclusions \label{sec:discuss}}

We have found the \hi\ reservoirs of the five blue UDG candidates around HCG07 and HCG25 that were identified by RT17 in optical images. We confirm that all of these objects are indeed UDGs with $R_e>1.5\,$kpc (Fig.~\ref{fig:spectra}).  Their systemic velocities imply that three of the five UDGs (UDG-B1, UDG-B4 and UDG-B5) are likely to be gravitationally bound to the HCG near which they project, that one (UDG-B3) is plausibly gravitationally bound to HCG25 and that one (UDG-B2) is in the background of HCG07 (Fig.~\ref{fig:vels}).

\begin{figure}
\epsscale{1.15}
\plotone{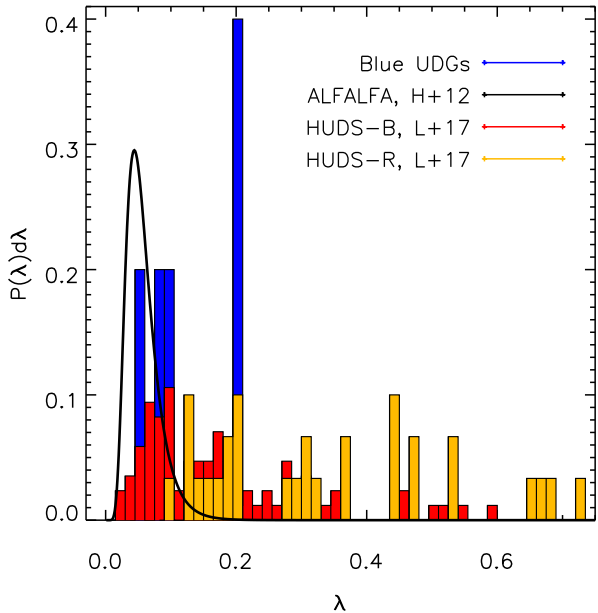}
\caption{Spin parameter estimator probability distribution $P(\lambda)d \lambda$ for the blue UDGs (blue histogram) as well as the HUDs-B (red histogram) and HUDs-R (yellow histogram) samples of \hi-bearing ultra diffuse sources from \citet{leisman17}. The best-fitting log-normal function to the ALFALFA 40\% sample analyzed by \citet{huang12b} is shown in black. All spins are computed using the estimator in Eq.~\ref{eq:lambda}.  \label{fig:lambda}}
\end{figure}

We use the integrated fluxes, line widths and distances measured from the \hi\ spectra together with the effective radii reported by RT17 to compute \hi\ masses and estimate dynamical masses and spin parameters for the blue UDGs (Table~\ref{tab:params}).  We find that their location in the \mhi --\mstar\ plane is broadly consistent with that of the gas-rich galaxy population probed by ALFALFA \citep{huang12b}, as well as with other gas-rich UDGs from the literature (Fig.~\ref{fig:MHIMstar}). The \hi\ masses and gas fractions of the blue UDGs correlate with their effective radii (Fig.~\ref{fig:SizeMass}).  The distribution of their spin parameter estimators appear to be more extreme than those of the broader ALFALFA population, but less so than the ``restricted'' HUDs-R sample of \hi -selected systems from \citep{leisman17}: the latter tend to have narrower velocity widths, higher gas fractions and larger spins than the blue UDGs (Figs.~\ref{fig:wfifty},~\ref{fig:MHIMstar}~and~\ref{fig:lambda}). 


  
Our measures of \Wfifty\ and \mhi\ as well as our estimates of \mdyn\ suggest that the blue UDGs are dwarf galaxies, reminiscent of the low-mass systems thought to make up the bulk of the red UDG population \citep{beasleytrujillo16,amorisco16,sifon17}. Combined with their elevated spins relative to the gas-rich galaxy population probed by ALFALFA, the blue UDGs share similar properties to the \hi -selected HUDs \citep{leisman17}.  Nonetheless, the blue UDG gas contents and spins are not as extreme as those of the HUDs-R sample;  these differences suggest that, as cautioned by \citet{leisman17}, \hi\ and optical searches may probe different subsets of the LSB galaxy population. We emphasize again that the dynamical masses and halo spin parameter estimators used here are only rough approximations. More reliable measures can be obtained when the \hi\ distribution is resolved \citep{hallenbeck14}; among the objects studied here, only UDG-B1 is a feasible target for higher-resolution interferometric follow-up than presented by \citet{borthakur10} and used here in order to obtain the required \hi\ maps.  
  

 If the blue UDGs that we confirm to have similar distances to HCG07 and HCG25 do indeed resemble the progenitors of the red UDGs therein, then what constraints do our observations place on models for their origin? The low masses of the blue UDGs as well as the correlation between their effective radii and their \hi\ masses and gas richnesses appear to be consistent with the dwarf galaxy simulations of \citet{dicintio17}. This model also predicts that, relative to gas-rich galaxies with similar \mstar, the blue UDGs should have an excess of \hi\ but similar halo spins. Instead, our observations suggest that the blue UDGs occupy a similar location in the \mhi --\mstar\ plane as the gas-rich galaxy population but have elevated halo spins, as predicted by other UDG formation models \citep{yozin15,amorisco16,rong17}. It is of course possible that several mechanisms contribute to the formation of UDGs, and that those proposed so far are effective for different subsets of the progenitor population. 
 
 
The data presented here highlight the potential of single-dish radio telescopes for measuring distances and masses for UDGs identified in the optical, although some care is required to avoid contamination from objects that fall within their relatively broad FWHMs (c.f.\ UDG-B1). By definition, stellar kinematics of UDGs are expensive to obtain: several hours on 8m-class telescopes are required to secure redshifts for individual objects \citep{vandokkum15,kadowaki17}, and building up sufficient sensitivity to measure a spectral width requires several nights \citep{vandokkum16,vandokkum17}. Such extreme measures are necessary to probe the internal kinematics of gas-poor objects, but the outlook is significantly better for relatively isolated gas-rich ones for which redshifts, \hi\ masses and velocity widths can sometimes be obtained in minutes \citep[][this work]{papastergis17}.  Single-dish follow-up is therefore a promising gateway to measuring the physical properties of large samples of UDGs.

\acknowledgments

We thank Luke Leisman for kindly sharing his data in advance of publication and Sanchayeeta Borthakur for 
providing her VLA datacube of HCG25. We also thank the referee for a careful consideration of this manuscript and patience with the re-submission process.
KS acknowledges support from the National Science and Engineering Research Council of Canada. 
The GBT is operated by the Green Bank Observatory, which is a facility of the National Science Foundation.
This research has made use of the NASA/IPAC Extragalactic Database (NED) which
is operated by the Jet Propulsion Laboratory, California Institute of Technology, 
under contract with the National Aeronautics and Space Administration.
Funding for the Sloan Digital Sky Survey IV has been provided by the Alfred P. Sloan Foundation, the U.S. Department of Energy Office of Science, and the Participating Institutions. SDSS-IV acknowledges
support and resources from the Center for High-Performance Computing at
the University of Utah. The SDSS web site is www.sdss.org.

%

\vspace{5mm}
\facilities{GBT (VEGAS)}





\bibliography{refs}

\begin{thebibliography}{}
\expandafter\ifx\csname natexlab\endcsname\relax\def\natexlab#1{#1}\fi
\providecommand{\url}[1]{\href{#1}{#1}}

\bibitem[{{Abraham} \& {van Dokkum}(2014)}]{abraham14}
{Abraham}, R.~G., \& {van Dokkum}, P.~G. 2014, \pasp, 126, 55

\bibitem[{{Amorisco} \& {Loeb}(2016)}]{amorisco16}
{Amorisco}, N.~C., \& {Loeb}, A. 2016, \mnras, 459, L51

\bibitem[{{Beasley} {et~al.}(2016){Beasley}, {Romanowsky}, {Pota}, {Navarro},
  {Martinez Delgado}, {Neyer}, \& {Deich}}]{beasley16}
{Beasley}, M.~A., {Romanowsky}, A.~J., {Pota}, V., {et~al.} 2016, \apjl, 819,
  L20

\bibitem[{{Beasley} \& {Trujillo}(2016)}]{beasleytrujillo16}
{Beasley}, M.~A., \& {Trujillo}, I. 2016, \apj, 830, 23

\bibitem[{{Bellazzini} {et~al.}(2017){Bellazzini}, {Belokurov}, {Magrini},
  {Fraternali}, {Testa}, {Beccari}, {Marchetti}, \& {Carini}}]{bellazzini17}
{Bellazzini}, M., {Belokurov}, V., {Magrini}, L., {et~al.} 2017, \mnras, 467,
  3751

\bibitem[{{Borthakur} {et~al.}(2010){Borthakur}, {Yun}, \&
  {Verdes-Montenegro}}]{borthakur10}
{Borthakur}, S., {Yun}, M.~S., \& {Verdes-Montenegro}, L. 2010, \apj, 710, 385

\bibitem[{{Bothun} {et~al.}(1991){Bothun}, {Impey}, \& {Malin}}]{bothun91}
{Bothun}, G.~D., {Impey}, C.~D., \& {Malin}, D.~F. 1991, \apj, 376, 404

\bibitem[{{Bottinelli} {et~al.}(1990){Bottinelli}, {Gouguenheim}, {Fouque}, \&
  {Paturel}}]{bottinelli90}
{Bottinelli}, L., {Gouguenheim}, L., {Fouque}, P., \& {Paturel}, G. 1990,
  \aaps, 82, 391

\bibitem[{{Bradford} {et~al.}(2015){Bradford}, {Geha}, \&
  {Blanton}}]{bradford15}
{Bradford}, J.~D., {Geha}, M.~C., \& {Blanton}, M.~R. 2015, \apj, 809, 146

\bibitem[{{Broeils} \& {Rhee}(1997)}]{broeils97}
{Broeils}, A.~H., \& {Rhee}, M.-H. 1997, \aap, 324, 877

\bibitem[{{Burkholder} {et~al.}(2001){Burkholder}, {Impey}, \&
  {Sprayberry}}]{burkholder01}
{Burkholder}, V., {Impey}, C., \& {Sprayberry}, D. 2001, \aj, 122, 2318

\bibitem[{{Catinella} {et~al.}(2006){Catinella}, {Giovanelli}, \&
  {Haynes}}]{catinella06}
{Catinella}, B., {Giovanelli}, R., \& {Haynes}, M.~P. 2006, \apj, 640, 751

\bibitem[{{Chang} {et~al.}(2015){Chang}, {van der Wel}, {da Cunha}, \&
  {Rix}}]{chang15}
{Chang}, Y.-Y., {van der Wel}, A., {da Cunha}, E., \& {Rix}, H.-W. 2015, \apjs,
  219, 8

\bibitem[{{Cook} {et~al.}(2014){Cook}, {Dale}, {Johnson}, {Van Zee}, {Lee},
  {Kennicutt}, {Calzetti}, {Staudaher}, \& {Engelbracht}}]{cook14}
{Cook}, D.~O., {Dale}, D.~A., {Johnson}, B.~D., {et~al.} 2014, \mnras, 445, 899

\bibitem[{{Di Cintio} {et~al.}(2017){Di Cintio}, {Brook}, {Dutton},
  {Macci{\`o}}, {Obreja}, \& {Dekel}}]{dicintio17}
{Di Cintio}, A., {Brook}, C.~B., {Dutton}, A.~A., {et~al.} 2017, \mnras, 466,
  L1

\bibitem[{{Giovanelli} \& {Haynes}(1988)}]{giov88}
{Giovanelli}, R., \& {Haynes}, M.~P. 1988, {Extragalactic neutral hydrogen},
  ed. K.~I. {Kellermann} \& G.~L. {Verschuur}, 522--562

\bibitem[{{Giovanelli} {et~al.}(2005){Giovanelli}, {Haynes}, {Kent},
  {Perillat}, {Saintonge}, {Brosch}, {Catinella}, {Hoffman}, {Stierwalt},
  {Spekkens}, {Lerner}, {Masters}, {Momjian}, {Rosenberg}, {Springob},
  {Boselli}, {Charmandaris}, {Darling}, {Davies}, {Garcia Lambas}, {Gavazzi},
  {Giovanardi}, {Hardy}, {Hunt}, {Iovino}, {Karachentsev}, {Karachentseva},
  {Koopmann}, {Marinoni}, {Minchin}, {Muller}, {Putman}, {Pantoja}, {Salzer},
  {Scodeggio}, {Skillman}, {Solanes}, {Valotto}, {van Driel}, \& {van
  Zee}}]{giov05}
{Giovanelli}, R., {Haynes}, M.~P., {Kent}, B.~R., {et~al.} 2005, \aj, 130, 2598

\bibitem[{{Hallenbeck} {et~al.}(2014){Hallenbeck}, {Huang}, {Spekkens},
  {Haynes}, {Giovanelli}, {Adams}, {Brinchmann}, {Chengalur}, {Hunt},
  {Masters}, \& {Saintonge}}]{hallenbeck14}
{Hallenbeck}, G., {Huang}, S., {Spekkens}, K., {et~al.} 2014, \aj, 148, 69

\bibitem[{{Haynes} {et~al.}(2011){Haynes}, {Giovanelli}, {Martin}, {Hess},
  {Saintonge}, {Adams}, {Hallenbeck}, {Hoffman}, {Huang}, {Kent}, {Koopmann},
  {Papastergis}, {Stierwalt}, {Balonek}, {Craig}, {Higdon}, {Kornreich},
  {Miller}, {O'Donoghue}, {Olowin}, {Rosenberg}, {Spekkens}, {Troischt}, \&
  {Wilcots}}]{haynes11}
{Haynes}, M.~P., {Giovanelli}, R., {Martin}, A.~M., {et~al.} 2011, \aj, 142,
  170

\bibitem[{{Hernandez} {et~al.}(2007){Hernandez}, {Park}, {Cervantes-Sodi}, \&
  {Choi}}]{hernandez07}
{Hernandez}, X., {Park}, C., {Cervantes-Sodi}, B., \& {Choi}, Y.-Y. 2007,
  \mnras, 375, 163

\bibitem[{{Hickson}(1982)}]{hickson82}
{Hickson}, P. 1982, \apj, 255, 382

\bibitem[{{Hickson} {et~al.}(1992){Hickson}, {Mendes de Oliveira}, {Huchra}, \&
  {Palumbo}}]{hickson92}
{Hickson}, P., {Mendes de Oliveira}, C., {Huchra}, J.~P., \& {Palumbo}, G.~G.
  1992, \apj, 399, 353

\bibitem[{{Huang} {et~al.}(2012){Huang}, {Haynes}, {Giovanelli}, \&
  {Brinchmann}}]{huang12b}
{Huang}, S., {Haynes}, M.~P., {Giovanelli}, R., \& {Brinchmann}, J. 2012, \apj,
  756, 113

\bibitem[{{Hunter} \& {Gallagher}(1985)}]{hunter85}
{Hunter}, D.~A., \& {Gallagher}, III, J.~S. 1985, \aj, 90, 1789

\bibitem[{{Impey} {et~al.}(1988){Impey}, {Bothun}, \& {Malin}}]{impey88}
{Impey}, C., {Bothun}, G., \& {Malin}, D. 1988, \apj, 330, 634

\bibitem[{{James} {et~al.}(2015){James}, {Koposov}, {Stark}, {Belokurov},
  {Pettini}, \& {Olszewski}}]{james15}
{James}, B.~L., {Koposov}, S., {Stark}, D.~P., {et~al.} 2015, \mnras, 448, 2687

\bibitem[{{Kadowaki} {et~al.}(2017){Kadowaki}, {Zaritsky}, \&
  {Donnerstein}}]{kadowaki17}
{Kadowaki}, J., {Zaritsky}, D., \& {Donnerstein}, R.~L. 2017, \apjl, 838, L21

\bibitem[{{Kalberla} \& {Kerp}(2009)}]{kalberla09}
{Kalberla}, P.~M.~W., \& {Kerp}, J. 2009, \araa, 47, 27

\bibitem[{{Koda} {et~al.}(2015){Koda}, {Yagi}, {Yamanoi}, \&
  {Komiyama}}]{koda15}
{Koda}, J., {Yagi}, M., {Yamanoi}, H., \& {Komiyama}, Y. 2015, \apjl, 807, L2

\bibitem[{{Konstantopoulos} {et~al.}(2010){Konstantopoulos}, {Gallagher},
  {Fedotov}, {Durrell}, {Heiderman}, {Elmegreen}, {Charlton}, {Hibbard},
  {Tzanavaris}, {Chandar}, {Johnson}, {Maybhate}, {Zabludoff}, {Gronwall},
  {Szathmary}, {Hornschemeier}, {English}, {Whitmore}, {Mendes de Oliveira}, \&
  {Mulchaey}}]{konstantopoulos10}
{Konstantopoulos}, I.~S., {Gallagher}, S.~C., {Fedotov}, K., {et~al.} 2010,
  \apj, 723, 197

\bibitem[{{Leisman} {et~al.}(2017){Leisman}, {Haynes}, {Janowiecki},
  {Hallenbeck}, {J{\'o}zsa}, {Giovanelli}, {Adams}, {Bernal Neira}, {Cannon},
  {Janesh}, {Rhode}, \& {Salzer}}]{leisman17}
{Leisman}, L., {Haynes}, M.~P., {Janowiecki}, S., {et~al.} 2017, ArXiv
  e-prints, arXiv:1703.05293

\bibitem[{{Licquia} \& {Newman}(2015)}]{licquia15}
{Licquia}, T.~C., \& {Newman}, J.~A. 2015, \apj, 806, 96

\bibitem[{{Mould} {et~al.}(2000){Mould}, {Huchra}, {Freedman}, {Kennicutt},
  {Ferrarese}, {Ford}, {Gibson}, {Graham}, {Hughes}, {Illingworth}, {Kelson},
  {Macri}, {Madore}, {Sakai}, {Sebo}, {Silbermann}, \& {Stetson}}]{mould00}
{Mould}, J.~R., {Huchra}, J.~P., {Freedman}, W.~L., {et~al.} 2000, \apj, 529,
  786

\bibitem[{{Mu{\~n}oz-Mateos} {et~al.}(2015){Mu{\~n}oz-Mateos}, {Sheth},
  {Regan}, {Kim}, {Laine}, {Erroz-Ferrer}, {Gil de Paz}, {Comeron}, {Hinz},
  {Laurikainen}, {Salo}, {Athanassoula}, {Bosma}, {Bouquin}, {Schinnerer},
  {Ho}, {Zaritsky}, {Gadotti}, {Madore}, {Holwerda}, {Men{\'e}ndez-Delmestre},
  {Knapen}, {Meidt}, {Querejeta}, {Mizusawa}, {Seibert}, {Laine}, \&
  {Courtois}}]{munoz15}
{Mu{\~n}oz-Mateos}, J.~C., {Sheth}, K., {Regan}, M., {et~al.} 2015, \apjs, 219,
  3

\bibitem[{{Munari} {et~al.}(2013){Munari}, {Biviano}, {Borgani}, {Murante}, \&
  {Fabjan}}]{munari13}
{Munari}, E., {Biviano}, A., {Borgani}, S., {Murante}, G., \& {Fabjan}, D.
  2013, \mnras, 430, 2638

\bibitem[{{O'Neil} \& {Bothun}(2000)}]{oneil00}
{O'Neil}, K., \& {Bothun}, G. 2000, \apj, 529, 811

\bibitem[{{Papastergis} {et~al.}(2017){Papastergis}, {Adams}, \&
  {Romanowsky}}]{papastergis17}
{Papastergis}, E., {Adams}, E.~A.~K., \& {Romanowsky}, A.~J. 2017, \aap, 601,
  L10

\bibitem[{{Pickering} {et~al.}(1997){Pickering}, {Impey}, {van Gorkom}, \&
  {Bothun}}]{pickering97}
{Pickering}, T.~E., {Impey}, C.~D., {van Gorkom}, J.~H., \& {Bothun}, G.~D.
  1997, \aj, 114, 1858

\bibitem[{{Rom{\'a}n} \& {Trujillo}(2017)}]{roman17}
{Rom{\'a}n}, J., \& {Trujillo}, I. 2017, \mnras, 468, 4039

\bibitem[{{Rong} {et~al.}(2017){Rong}, {Guo}, {Gao}, {Liao}, {Xie}, {Puzia},
  {Sun}, \& {Pan}}]{rong17}
{Rong}, Y., {Guo}, Q., {Gao}, L., {et~al.} 2017, \mnras, 470, 4231

\bibitem[{{Salucci} {et~al.}(2007){Salucci}, {Lapi}, {Tonini}, {Gentile},
  {Yegorova}, \& {Klein}}]{salucci07}
{Salucci}, P., {Lapi}, A., {Tonini}, C., {et~al.} 2007, \mnras, 378, 41

\bibitem[{{SDSS Collaboration} {et~al.}(2016){SDSS Collaboration}, {Albareti},
  {Allende Prieto}, {Almeida}, {Anders}, {Anderson}, {Andrews},
  {Aragon-Salamanca}, {Argudo-Fernandez}, {Armengaud}, \& et~al.}]{dr13}
{SDSS Collaboration}, {Albareti}, F.~D., {Allende Prieto}, C., {et~al.} 2016,
  ArXiv e-prints, arXiv:1608.02013

\bibitem[{{Shi} {et~al.}(2017){Shi}, {Zheng}, {Zhao}, {Pan}, {Li}, {Zou},
  {Zhou}, {Guo}, {An}, \& {Li}}]{shi17}
{Shi}, D., {Zheng}, X., {Zhao}, H., {et~al.} 2017, ArXiv e-prints,
  arXiv:1708.00013

\bibitem[{{Sif{\'o}n} {et~al.}(2017){Sif{\'o}n}, {van der Burg}, {Hoekstra},
  {Muzzin}, \& {Herbonnet}}]{sifon17}
{Sif{\'o}n}, C., {van der Burg}, R.~F.~J., {Hoekstra}, H., {Muzzin}, A., \&
  {Herbonnet}, R. 2017, ArXiv e-prints, arXiv:1704.07847

\bibitem[{{Springob} {et~al.}(2005){Springob}, {Haynes}, {Giovanelli}, \&
  {Kent}}]{springob05}
{Springob}, C.~M., {Haynes}, M.~P., {Giovanelli}, R., \& {Kent}, B.~R. 2005,
  \apjs, 160, 149

\bibitem[{{Trujillo} \& {Fliri}(2016)}]{trujillo16}
{Trujillo}, I., \& {Fliri}, J. 2016, \apj, 823, 123

\bibitem[{{Trujillo} {et~al.}(2017){Trujillo}, {Roman}, {Filho}, \&
  {S{\'a}nchez Almeida}}]{trujillo17}
{Trujillo}, I., {Roman}, J., {Filho}, M., \& {S{\'a}nchez Almeida}, J. 2017,
  \apj, 836, 191

\bibitem[{{van der Burg} {et~al.}(2016){van der Burg}, {Muzzin}, \&
  {Hoekstra}}]{vanderburg16}
{van der Burg}, R.~F.~J., {Muzzin}, A., \& {Hoekstra}, H. 2016, \aap, 590, A20

\bibitem[{{van der Burg} {et~al.}(2017){van der Burg}, {Hoekstra}, {Muzzin},
  {Sifon}, {Viola}, {Bremer}, {Brough}, {Driver}, {Erben}, {Heymans},
  {Hildebrandt}, {Holwerda}, {Klaes}, {Kuijken}, {McGee}, {Nakajima},
  {Napolitano}, {Norberg}, {Taylor}, \& {Valentijn}}]{vanderburg17}
{van der Burg}, R.~F.~J., {Hoekstra}, H., {Muzzin}, A., {et~al.} 2017, ArXiv
  e-prints, arXiv:1706.02704

\bibitem[{{van Dokkum} {et~al.}(2016){van Dokkum}, {Abraham}, {Brodie},
  {Conroy}, {Danieli}, {Merritt}, {Mowla}, {Romanowsky}, \&
  {Zhang}}]{vandokkum16}
{van Dokkum}, P., {Abraham}, R., {Brodie}, J., {et~al.} 2016, \apjl, 828, L6

\bibitem[{{van Dokkum} {et~al.}(2017){van Dokkum}, {Abraham}, {Romanowsky},
  {Brodie}, {Conroy}, {Danieli}, {Lokhorst}, {Merritt}, {Mowla}, \&
  {Zhang}}]{vandokkum17}
{van Dokkum}, P., {Abraham}, R., {Romanowsky}, A.~J., {et~al.} 2017, \apjl,
  844, L11

\bibitem[{{van Dokkum} {et~al.}(2015){van Dokkum}, {Abraham}, {Merritt},
  {Zhang}, {Geha}, \& {Conroy}}]{vandokkum15}
{van Dokkum}, P.~G., {Abraham}, R., {Merritt}, A., {et~al.} 2015, \apjl, 798,
  L45

\bibitem[{{Yagi} {et~al.}(2016){Yagi}, {Koda}, {Komiyama}, \&
  {Yamanoi}}]{yagi16}
{Yagi}, M., {Koda}, J., {Komiyama}, Y., \& {Yamanoi}, H. 2016, \apjs, 225, 11

\bibitem[{{Yozin} \& {Bekki}(2015)}]{yozin15}
{Yozin}, C., \& {Bekki}, K. 2015, \mnras, 452, 937

\bibitem[{{Zaritsky}(2017)}]{zaritsky17}
{Zaritsky}, D. 2017, \mnras, 464, L110

\end{thebibliography}






\end{document}